\documentstyle[12pt]{article}
\makeatletter
\def\fmslash{\@ifnextchar[{\fmsl@sh}{\fmsl@sh[0mu]}}
\def\fmsl@sh[#1]#2{%
  \mathchoice
    {\@fmsl@sh\displaystyle{#1}{#2}}%
    {\@fmsl@sh\textstyle{#1}{#2}}%
    {\@fmsl@sh\scriptstyle{#1}{#2}}%
    {\@fmsl@sh\scriptscriptstyle{#1}{#2}}}
\def\@fmsl@sh#1#2#3{\m@th\ooalign{$\hfil#1\mkern#2/\hfil$\crcr$#1#3$}}
\makeatother
\input{epsf}
\begin{document}
%---------------- TTP Titlepage <---------------------------
\thispagestyle{empty}
\begin{titlepage}
\hskip -3cm
\begin{flushright}
{\bf TTP 95--06} \\
hep-ph/9503272
\end{flushright}
\vspace{1cm}
\begin{center}
{\Large\bf INCLUSIVE DECAYS \vspace*{3mm} \\
           OF HEAVY FLAVOURS}
\end{center}
\vspace{0.8cm}
\begin{center}
{\sc Thomas Mannel}  \vspace*{2mm} \\
{\sl Institut f\"{u}r Theoretische Teilchenphysik, University of Karlsruhe \\
     Kaiserstr. 12, D -- 76128 Karlsruhe, Germany.}
\end{center}
\vfill
\begin{abstract}
\noindent
Recent progress in the theoretical description of inclusive heavy flavour
decays is reviewed. After an outline of the theoretical methods applications to
total decay rates and semileptonic decay spectra are presented.
\end{abstract}
\vfill
\begin{center}{\it Contribution to the workshop on ``Heavy Quark Physics'',\\
                   December 13 -- 16, 1994, Bad Honnef, Germany.}
\end{center}
\vfill
\begin{flushleft}
{\bf TTP 95--06} \\
March 1995
\end{flushleft}
\end{titlepage}
%---------------- END TTP Titlepage <---------------------------
\newpage
\section{Introduction}
\noindent
Decays of heavy flavours play an important role in the determination of the
not yet well explored CKM sektor of the standard model. The weak processes
that need to be studied involve transitions among quarks of different flavours;
however, to connect the observed transitions among hadrons to the underlying
quark processes one has to deal with the bound state problem of strong
interactions.

For the decays of heavy flavours one may take advantage of the fact that the
mass $m_Q$ of the heavy quark is large compared to the scale $\bar\Lambda$,
which is determined by the light degrees of freedom and thus is of the order
of $\Lambda_{QCD}$. The amplitudes or the transition rates for the decays of
heavy flavoured hadrons are evaluated as an expansion in powers of
$\bar\Lambda / m_Q$, the leading term of which corresponds to an infinitely
heavy, static quark \cite{shifman,isgur/wise,reviews}.

The $1/m_Q$ expansion has been formulated in the laguage of effective field
theory, the so called Heavy Quark Effective Theory (HQET)
\cite{isgur/wise,reviews}.
In this framework
one may systematically access the mass dependence of any matrix element
involving heavy hadron states and heavy quark fields. Furthermore, in the
infinite mass limit two additional symmetries appear \cite{isgur/wise},
which play an important
role in the context of exclusive heavy flavour decays.

In the past few years methods have been developed to apply the $1/m_Q$
expansion also to inclusive decays of heavy hadrons
\cite{russnl,CCGincsl,Bigiincsl,Blokincsl,mwincsl,tmincsl,FLS93}, and the
purpose of
this talk is to give a short review of these developments. It is impossible
to cover all topics in the field of inclusive decays due to length and time
restrictions and consequently this review can cover only some selected
topics.

For inclusive decays a $1/m_Q$ expansion is obtained for the
rates by an approach similar to the one known from deep inelastic
scattering. The first step consists of an Operator Product Expansion (OPE)
which yields an infinite sum of operators with increasing dimension.
The dimensions of the operators are compensated by inverse powers
of a large scale, which is in general of the order of the heavy mass
scale. The decay probability is then given as forward matrix elements
of these operators between the state of the decaying heavy hadron;
these matrix elements still have a mass dependence, which then may
be extracted in terms of a $1/m_Q$ expansion using HQET as for exclusive
decays.

In section 2 we give a short description of the OPE approach to inclusive
decays. Section 3 summarizes the results for the total rates, including the
first non-trivial non-perturbative corrections. In section 3 the method
is applied to the charged lepton energy spectrum in inclusive semileptonic
$B$ decays. It turns out that the endpoint region cannot be described by a
$1/m_Q$ expansion; rather a partial resummation of the $1/m_Q$ expansion
is required, which is closely analogous to the leading twist term in
deep inelastic scattering \cite{NeubertShape,BiMotion,ManNeu,NeubertBsg}.
This is summarized in section 4. Finally, we
conclude and point out a few open questions which are currently under
study.

\section{Operator Product Expansion}
\noindent
The effective Hamiltonian for a decay of a heavy
(down-type) quark is in general linear in the decaying heavy
flavoured quark
\begin{equation} \label{heff}
{\cal H}_{eff} = \bar{Q} R
\end{equation}
where the operator $R$ describes the decay products.
In the following we shall consider semileptonic decays, for which
\begin{equation} \label{rsl}
R_{sl} = \frac{G_F}{\sqrt{2}} V_{Qq} \,\, \gamma_\mu (1-\gamma_5) q
\,\, (\bar{\nu}_\ell \gamma^\mu (1-\gamma_5) \ell)  ,
\end{equation}
where $q$ is an up-type quark ($c$ or $u$, since we shall consider $b$
decays). Similarly, for nonleptonic decays the Cabbibo allowed contribution
corresponds to
\begin{eqnarray}
R_{nl} &=& \frac{G_F}{2\sqrt{2}} V_{Qq} V^*_{q'q''} \left[ (C_+ (m_b) + C_-
(m_b))
         \gamma_\mu (1-\gamma_5) q  \right. (\bar{q}' \gamma^\mu (1-\gamma_5)
q'')
       \\
       && \qquad \qquad \qquad + (C_+ (m_b) - C_- (m_b))
         \gamma_\mu (1-\gamma_5) q''
       \left. (\bar{q}' \gamma^\mu (1-\gamma_5) q )\right],
\nonumber
\end{eqnarray}
where $q'$ ($q''$) is a  down-type (up-type) quark and $V_{Qq}$ the
corresponding
CKM matrix element. The coefficients $C_\pm (m_b)$
are the QCD corrections obtained from the renormalization group running between
$M_W$ and $m_b$; in leading logarithmic approximation these coefficients are
\cite{C12}
\begin{equation} \label{cpm}
C_{\pm} (m_b) =[\frac{\alpha _S(M_W^2)}{\alpha _S(m_b^2)}]^{\gamma_{\pm}},
\mbox{ with } \gamma _+=\frac{6}{33-2N_f}=-\frac{1}{2}\gamma _- ,
\end{equation}
where $\alpha_s (\mu)$ is the onle-loop expression for the running coupling
coupling constant of QCD.

Finally, for  radiative rare decays we have
\begin{equation}
R_{rare} = \frac{G_F}{\sqrt{2}} V_{tb} V^*_{ts} C_7 (m_b) \frac{e}{16 \pi^2}
            m_b \sigma_{\mu \nu} (1+\gamma_5) s F^{\mu \nu} ,
\end{equation}
where $C_7 (m_b) $ is again a coefficient obtained from running between $M_W$
and $m_b$. Its value is $C_7 (m_b) \sim 0.3$, the corresponding analytical
expression may be found in \cite{C7}.

The inclusive decay rate for a heavy hadron $H$ containing the quark $Q$
may be related to a forward matrix element by
\begin{eqnarray}
\Gamma  & \propto &   \sum_X (2 \pi)^4 \delta^4 (P_H - P_X )
| \langle X | {\cal H}_{eff} | H(v) \rangle |^2
\label{inclusive} \\
&=& \int d^4 x \,  \langle H(v) |{\cal H}_{eff} (x)
            {\cal H}_{eff}^\dagger (0) | H(v) \rangle
\nonumber \\ \nonumber
&=&  2 \mbox{ Im}
\int d^4 x \,  \langle H(v) |T \{ {\cal H}_{eff} (x)
            {\cal H}_{eff}^\dagger (0) \} | H(v) \rangle ,
\end{eqnarray}
where $| X \rangle $ is the final state which is summed over to
obtain the inclusive rate.

The matrix element appearing in (\ref{inclusive}) contains a
large scale, namely the mass of the heavy quark. The first step
towards a $1/m_Q$ expansion is to make this large scale explicit.
This may be  done by a phase redefinition which leads to
\begin{equation}
\Gamma  \propto
2 \mbox{ Im}
\int d^4 x  \, e^{-im_Q vx}
\langle H(v) |T \{ \widetilde{{\cal H}}_{eff} (x)
   \widetilde{{\cal H}}_{eff} ^\dagger (0) \} | H(v) \rangle ,
\end{equation}
where
\begin{equation} \label{phaseredef}
\widetilde{{\cal H}}_{eff}
= \bar{Q}_v R \qquad Q_v = e^{-im_Q vx} Q .
\end{equation}
This relation exhibits the similarity
between the cross-section calculation in deep inelastic scattering
and the present approach to total rates. In deep inelastic scattering
there appears a large scale which is the momentum transfer to the
leptons, while here the mass of the heavy quark appears as a large scale.

The next step is to perform an operator product expansion of the product
of the two Hamitonians. After the phase redefinition
the remaining matrix element does not involve large momenta of the order
of the heavy quark mass any more and hence a short-distance expansion
becomes useful, if the mass $m_Q$ is large compared to the scale
$\bar\Lambda$ determining the matrix element.
The next step is thus to perform an operator-product
expansion, which has the general form
\begin{eqnarray}
&& \int d^4 x  \, e^{im_Q vx}
\langle H(v) |T \{ \widetilde{{\cal H}}_{eff} (x)
\widetilde{{\cal H}}_{eff}^\dagger (0) \}| H(v) \rangle \\ \nonumber
&& \qquad \qquad = \sum_{n=0}^\infty  \left(\frac{1}{2m_Q}\right)^n
     \hat{C}_{n+3} (\mu) \, \langle H(v) |{\cal O}_{n+3}| H(v) \rangle_\mu ,
\end{eqnarray}
where ${\cal O}_n$ are operators of dimension $n$, with their
matrix elements renormalized at scale
$\mu$, and $\hat{C}_n$ are the corresponding Wilson coefficients.
These coefficients encode the short distance physics related to the
heavy quark mass scale and may be calculated in perturbation theory.
All long distance contributions connected to the hadronic scale
$\bar\Lambda$ are contained in the matrix elements of the operators
${\cal O}_{n+3}$.

Still the matrix elements of ${\cal O}_{n+3}$ are not independent of the
heavy quark mass scale, but this mass dependence may be expanded in powers
of $1/m_Q$ by means of heavy quark effective theory. This is achieved
by expanding the heavy quark fields appearing in the operators ${\cal O}_n$
as well as the states by including the corrections to the Lagrangian
as time-ordered products. In this way the
mass dependence of the total decay rate may be accessed completely within
an expansion in $1/m_Q$.

The lowest-order term of the operator product expansion are the
dimension-3 operators. Due to Lorentz invariance and parity there
are only two combinations which may appear, namely
$\bar  Q_v \fmslash{v} Q_v$ or $\bar Q_v Q_v$. Note that the $Q_v$
operators differ from the full QCD operators only by a phase redefinition,
and hence $\bar  Q_v \fmslash{v} Q_v = \bar  Q \fmslash{v} Q$ and
$\bar Q_v Q_v = \bar Q Q$.  The first combination
is proportional to the $Q$-number current $\bar  Q \gamma_\mu Q$,
which is normalized even in full QCD,
while the second differs from the first one only by terms of order
$1/m_Q^2$
\begin{equation}
\bar Q_v Q_v = v_\mu \bar Q_v \gamma_\mu Q_v
+ \frac{1}{2 m_Q^2}
\bar{h}_v  \left[ (iD)^2 - (ivD)^2 +
\frac{i}{2} \sigma_{\mu \nu} G^{\mu \nu} \right] h_v
+ {\cal O} (1/m_Q^3) ,
\end{equation}
where $G_{\mu \nu}$ is the gluon field strength.

Thus the matrix elements of the dimension-3 contribution
is known to be normalized; in the standard normalization of the
states this implies
\begin{equation}
\langle H(v) |{\cal O}_3| H(v) \rangle =
\langle H(v) |\bar Q_v \fmslash{v} Q_v | H(v) \rangle = 2 m_H ,
\end{equation}
where $m_H$ is the mass of the heavy hadron.
To lowest order in the heavy mass expansion we may furthermore
replace $m_H = m_Q$ and hence we may evaluate the leading term in
the $1/m_Q$ expansion without any hadronic uncertainty. Generically
the dimension-3 contribution yields the free quark decay rate.
This has been previously used as a model for inclusive decays, but
now it turns out to be the first term in a systematic $1/m_Q$ expansion
of total rates.

A dimension-4 operators contains an additional covariant derivative,
and thus one has matrix elements of the type
\begin{equation}
\langle H(v) |{\cal O}_4| H(v) \rangle \propto
\langle H(v) |\bar Q_v \Gamma D_\mu Q_v | H(v) \rangle = {\cal A}_\Gamma v_\mu
{}.
\end{equation}
Since the equations of motion apply for this tree level matrix element,
one finds that the constant ${\cal A}_\Gamma$ has to vanish, and thus
there are no dimension-4 contributions. This statement is completely
equivalent to Lukes theorem \cite{Luke}, since we are considering
a forward matrix element, i.e. a matrix element at zero recoil \cite{Mzerec}.

The first non-trivial non-perturbative contribution comes from
dimension-5 operators and are of order $1/m_Q^2$.
For mesonic decays there are only the two parameters $\lambda_1$
and $\lambda_2$ corresponding to
matrix elements involving higher order terms that appear in the effective
theory Lagrangian
\begin{eqnarray}
\lambda_1 &=& \frac{\langle H (v) | \bar{h}_v  (iD)^2  h_v | H (v) \rangle}
                   {2 M_H}
\label{lam1} \\
\lambda_2  &=& \frac{\langle H (v) | \bar{h}_v \sigma_{\mu \nu} iD^\mu iD^\nu
h_v
                                   | H (v) \rangle}
                    {2 M_H} ,
\label{lam2}
\end{eqnarray}
where the normalization of the states is chosen to be
$\langle H (v) | \bar{h}_v h_v | H (v) \rangle = 2 M_H$, where $M_H$ is the
mass of the heavy meson in the static limit.
These parameters may be interpreted as the expectation value of
the kinetic energy  of the heavy quark and its energy
due to the chromomagnetic moment of the heavy quark
inside the heavy meson respectively.

The parameter $\lambda_2$ is easy to access, since it is related to
the mass splitting between $H(v)$
and $H^* (v , \epsilon)$. From the $B$-meson system we obtain
\begin{equation}
\lambda_2 (m_b) = \frac{1}{4} (M_{H^*} - M_H) = 0.12 \mbox{ GeV}^2;
\end{equation}
from the charm system the same value is obtained.  This shows that indeed
the spin-symmetry partners are degenerate in the infinite mass limit and
the splitting between them scales as $1/m_Q$.

The parameter $\lambda_1$ appearing is not simply related to the hadron
spectrum;
from the definition of $\lambda_1$ one is led to assume
$\lambda_1 < 0$; a more restrictive inequality
\begin{equation}
-\lambda_1 > 3 \lambda_2
\end{equation}
has been derived in a quantum mechanical framework in \cite{BiMotion}
and using heavy-flavour sum rules \cite{Bisumrule}.
Furthermore, there exists also a QCD sum rule estimate \cite{BBsumrule}
for this parameter:
\begin{equation}
\lambda_1  = - 0.52 \pm 0.12 \mbox{ GeV}^2 .
\end{equation}

\section{Total Decay Rates}
\noindent
In this section we collect the results for the total rates
including the first non-trivial non-perturbative correction.

Inserting $R_{sl}$ as given in (\ref{rsl}) one obtains for the total
inclusive semileptonic decay rate $B \to X_c \ell \nu$
\begin{equation} \label{bcsl}
\Gamma_{B \to X_c \ell \nu}
 = \frac{G_F^2 m_b^5}{192\pi^3} |V_{cb}|^2
      \left[
      \left(1+\frac{\lambda_1}{2m_c^2}\right) f_1 \left(\frac{m_c}{m_b}\right)
      -\frac{9\lambda_2}{2m_c^2} f_2 \left(\frac{m_c}{m_b}\right)
      \right] ,
\end{equation}
where the two $f_j$ are phase-space functions
\begin{eqnarray} \label{f1}
f_1(x) &=& 1-8x^2+8x^6-x^8-24x^4\log x ,
\\  \nonumber
f_2(x) &=& 1-\frac{8}{3}x^2-8x^4+8x^6+\frac{5}{3}x^8+8x^4\log x  .
\end{eqnarray}
The result for $B \to X_u \ell \nu_\ell$ is obtained from (\ref{bcsl})
as the limit $m_c \to 0$ and the replacement $V_{cb} \to V_{ub}$
\begin{equation} \label{busl}
\Gamma_{B \to X_u \ell \nu} = \frac{G_F^2 m_b^5}{192\pi^3} |V_{ub}|^2
 \left[
      1+\frac{\lambda_1 - 9 \lambda_2}{2m_b^2}
      \right]   .
\end{equation}
As was discussed above, the leading non-perturbative
corrections in (\ref{bcsl}) and (\ref{busl}) are parametrized
by $\lambda_1$ and $\lambda_2$. Estimates for these parameters
have been discussed in section 2; in order to
estimate the total effect of the non-perturbative effects we
insert a range of values $-0.3 > \lambda_1 >  -0.6$ GeV${}^2$; from this
we obtain
\begin{equation}
\frac{\lambda_1 - 9 \lambda_2}{2m_b^2} \sim -(3 \cdots 4) \%
\end{equation}
This means that the non-perturbative contributions are small,
in particular compared to the perturbative ones, which have been
calculated some time ago \cite{QCDradcorr,ACCMM}. For the decay
$B \to X_u \ell \bar\nu_\ell$ the lowest order QCD corrections
are given by
\begin{equation}
\Gamma_{B \to X_u \ell \bar{\nu}_\ell} =  \frac{G_F^2 m_b^5}{192\pi^3}
       |V_{ub}|^2 \left[ 1 + \frac{2 \alpha}{3 \pi}
                    \left(\frac{25}{4}-\pi^2\right) \right]
= 0.85 |V_{ub}|^2 \Gamma_b,
\end{equation}
and thus the typical size of QCD radiative corrections is of the order of
ten to twenty percent.

Similarly, one obtains the result for non-leptonic decays as
\begin{equation}
\Gamma_{X_c} = 3 \frac{G_F^2 m_b^5}{192\pi^3}
\left\{ A_1 f_1 \left(\frac{m_c}{m_b} \right)
\left[ 1 + \frac{1}{2 m_b^2} (\lambda_1 - 9 \lambda_2) \right]
        - 48 A_2 f_3 \left(\frac{m_c}{m_b}\right)
          \frac{1}{2m_b^2} \lambda_2 \right\} ,
\end{equation}
where the coefficients $A_i$ are expressed as combination of the Wilson
coefficents
$C_\pm (m_b)$ given in (\ref{cpm})
\begin{equation}
A_1 = \frac{1}{3} [C_-^2 (m_b) + 2 C_+^2 (m_b)] , \qquad
A_2 = \frac{1}{6} [C_+^2 (m_b) - C_-^2 (m_b)] ,
\end{equation}
and $ f_3 (x) = (1-x^2)^3 $ is another phase space function.
Again the non-perturbative corrections turn out to be small, in the
region of a few percent compared to the leading term, and the perturbative
corrections turn out to be much larger than this.

Finally, for the rare decay $B \to X_s \gamma$ one may as well calculate
the non-perturbative contribution in terms of $\lambda_1$ and $\lambda_2$.
One obtains
\begin{equation}
\Gamma_{B\to X_s\gamma}
     =  \frac{\alpha G_F^2}{16\pi^4}
     m_b^5 |V_{ts} V_{td}^*|^2 |C_7(m_b)|^2
    \left[1+\frac{1}{2m_b^2}
    \left(\lambda_1-9\lambda_2 \right)\right],
\end{equation}
and the relative size of the nonperturbative corrections is the
same as in the $B \to X_u \ell \bar\nu_\ell$ decays.

Typically the non-pertubative corrections are much smaller than the radiative
corrections. The only exception is the endpoint region of lepton energy spectra
which receives both large perturbative as well as nonperturbative corrections.
However, this is only a small region in phase space and the corrections to the
total rates remain moderate.

\section{Lepton Energy Spectra}
\noindent
The method of the operator-product expansion may also  be used to
obtain the non-perturbative corrections to the charged lepton energy
spectrum. In this case the operator product expansion is
applied not to the full effective Hamiltonian, but rather only to the
hadronic currents. The rate is written as a product of the hadronic and
leptonic tensor
\begin{equation}
d \Gamma = \frac{G_F^2}{4 m_B} | V_{Qq} |^2 W_{\mu \nu}
\Lambda^{\mu \nu} d(PS) ,
\end{equation}
where $d(PS)$ is the phase-space differential.
The short-distance expansion is then performed for
the two currents appearing in
the hadronic tensor. Redefining the phase of the heavy-quark fields as in
(\ref{phaseredef}) one finds that the momentum transfer variable
relevant for the short-distance expansion is $m_Q v - q$, where
$q$ is the momentum transfer to the leptons.

The structure of the expansion for the spectrum is identical to the one
of the total rate. The contribution of the dimension-3 operators
yields the free-quark decay spectrum, there are no contributions from
dimension-4 operators, and the $1/m_b^2$ corrections are parametrized
in terms of $\lambda_1$ and $\lambda_2$. Calculating the spectrum for
$B \to X_c \ell \nu$ yields \cite{Bigiincsl,Blokincsl,mwincsl,tmincsl}
\begin{eqnarray}
\frac{d \Gamma}{dy} &=& \frac{G_F^2\,|\,V_{cb}|^2\,m_b^5}{192\pi^3}
\Theta(1-y-\rho) y^2 \left[\left\{ 3 (1-\rho) (1-R^2) - 2 y (1-R^3) \right\}
\right. \\
&& + \frac{\lambda_1}{[ m_b(1-y)]^2} (3 R^2-4 R^3)
  - \frac{\lambda_1}{m_b^2(1-y)} (R^2-2 R^3)
\nonumber \\
&& - \frac{3\lambda_2}{m_b^2(1-y)} (2 R+3 R^2-5 R^3)
   + \frac{\lambda_1}{3 m_b^2} [ 5y - 2(3-\rho) R^2 + 4 R^3 ]
\nonumber\\
&& \left. + \frac{\lambda_2}{m_b^2} [ (6+5y) - 12 R
    - (9-5\rho) R^2 + 10 R^3 ] \right]
    + {\cal O}\left[ (\Lambda/[m_b(1-y)])^3 \right] ,
\nonumber
\end{eqnarray}
where we have defined
\begin{equation}
\rho = \left(\frac{m_c}{m_b}\right)^2 , \quad  R = \frac{\rho}{1-y}
\end{equation}
and
\begin{equation}
y = 2 E_\ell / m_b
\end{equation}
is the rescaled energy of the charged lepton.

This expression is somewhat complicated, but it simplifies
for the decay $B \to X_u \ell \nu$ since then the mass of the quark
in the final state may be neglected. One finds
\begin{eqnarray} \label{buspec}
\frac{d\Gamma}{dy} &=&  \frac{G_F^2\,|\,V_{ub}|^2\,m_b^5}{192\pi^3}
        \left[ \left( 2y^2 (3-2y)
        + \frac{10y^2}{3} \frac{\lambda_1}{m_b^2}
        + 2y(6+5y) \frac{\lambda_2}{m_b^2} \right) \Theta(1-y) \right.
\nonumber \\
&&  \vphantom{\frac{G_F^2\,|\,V_{qb}|^2\,m_b^5}{192\pi^3}}
\left. - \frac{\lambda_1 + 33 \lambda_2}{3m_b^2} \delta (1-y)
  -  \frac{\lambda_1}{3m_b^2} \delta ' (1-y) \right] .
\end{eqnarray}

\begin{figure}
   \epsfysize=8cm
   \centerline{\epsffile{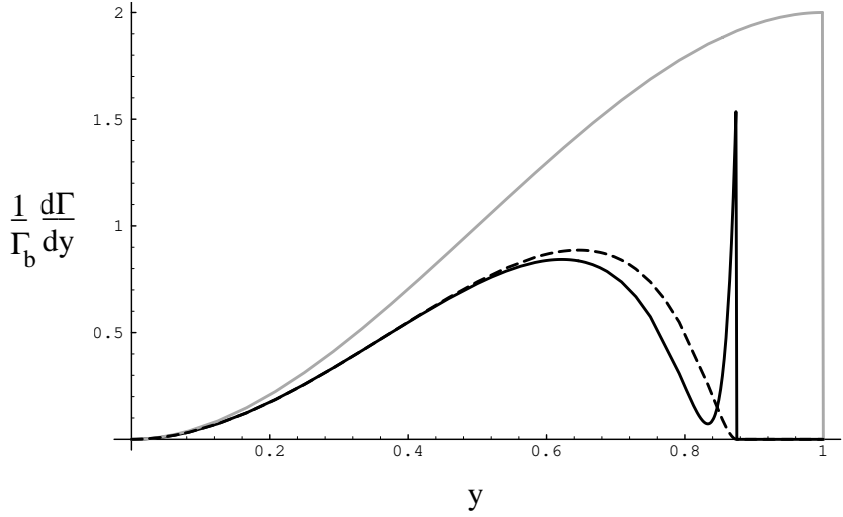}}
   \vskip -1.5cm
   \caption{The electron spectrum for free quark
$b\rightarrow c$ decay (dashed line), free quark $b\rightarrow u$ decay
(grey line), and  $B\rightarrow X_c e \bar\nu_e$ decay including
$1/m_b^2$ corrections (solid line) with $\lambda_1 = - 0.5$ GeV${}^2$
and $\lambda_2 = 0.12$ GeV${}^2$. The figure is from
\protect\cite{mwincsl}.}
\label{fig5}
\end{figure}

Figure~\ref{fig5} shows the distributions for inclusive semileptonic
decays of $B$ mesons. The spectrum close to the endpoint, where the
lepton energy becomes maximal, exhibits a sharp spike as $y \to y_{max}$.
In this region we have
\begin{equation}
\frac{d\Gamma}{dy}  \propto \Theta(1-y-\rho)
\left[ 2 + \frac{\lambda_1}{(m_Q (1-y))^2}
\left(\frac{\rho}{1-\rho} \right)^2
\left\{ 3 - 4 \left(\frac{\rho}{1-\rho} \right) \right\}  \right] ,
\end{equation}
which behaves like
$\delta$-functions and its derivatives as $\rho \to 0$, which can be
seen in (\ref{buspec}). This behaviour indicates a breakdown of the operator
product expansion close to the endpoint, since for the spectra the
expansion parameter is not $1/m_Q$, but rather $1/(m_Q - qv)$, which
becomes $1/(m_Q [1-y])$ after the integration over the neutrino momentum.
In order to obtain a description of the endpoint region, one has to perform
some resummation of the operator product expansion.

\section{Resummation in the Endpoint Region}
Very close to the endpoint of the inclusive semileptonic decay spectra
only a few resonances contribute. In this resonance region one cannot
expect to have a good description of the spectrum using
an approach based on parton-hadron duality; here a sum over a few
resonances will be appropriate.

In the variable $y$ the size of this resonance region is, however, of the
order of $(\bar\Lambda / m_Q)^2$ and thus small. In a larger region of
the order $\bar\Lambda / m_Q $, which we shall call the endpoint region,
many resonances contribute and one may hope to describe the spectrum
in this region using parton-hadron duality.

It has been argued in \cite{NeubertShape} that the $\delta$-function-like
singularities appearing in (\ref{buspec}) may be reinterpreted as
the expansion of a non-perturbative function describing the spectrum in the
endpoint region. Keeping only the singular
terms of (\ref{buspec}) we write
\begin{equation} \label{shape}
\frac{1}{\Gamma_b} \frac{d\Gamma}{dy}  = 2y^2 (3-2y) S(y) ,
\end{equation}
where
\begin{equation} \label{momexp}
S(y) = \Theta (1-y) + \sum_{n=0}^{\infty} a_n \delta^{(n)} (1-y)
\end{equation}
is a non-perturbative function given in terms of the moments $a_n$ of the
spectrum, taken over the endpoint region. These moments themselves have an
expansion in $1/m_Q$ such that $a_n \sim 1/m_Q^{n+1}$,
and we shall consider only the leading term in
the expansion of the moments, corresponding to the most singular
contribution to the endpoint region.

Comparing (\ref{buspec}) with
(\ref{shape}) and (\ref{momexp}) one obtains that
\begin{eqnarray}
a_0 &=& \int dy (S(y) - \Theta (1-y)) = {\cal O} (1/m_Q^2) , \\ \label{a1}
a_1 &=& \int y  (S(y) - \Theta (1-y)) = -\frac{\lambda_1}{3 m_Q^2} ,
\end{eqnarray}
where the integral extends over the endpoint region.

The non-perturbative function implements a resummation of the
most singular terms contributing to the endpoint and, in the language
of deep inelastic scattering, corresponds to the
leading twist contribution. A similar approach based on the parton model
will be described in a separate talk at this conference \cite{Paschos}.

This
resummation has been studied in QCD \cite{BiMotion,ManNeu} and the
function $S(y)$ may be related to the distribution of the light cone
component of the heavy quark residual momentum
inside the heavy meson. The latter is a fundamental
function for inclusive heavy-to-light transitions, which has been
defined in \cite{BiMotion}
\begin{equation}\label{fdef}
   f(k_+) = \frac{1}{2m_B}
   \langle H(v)|\,\bar h_v\,\delta(k_+-i D_+)\,h_v\,
   |H(v)\rangle ,
\end{equation}
where $k_+ = k_0 + k_3$ is the positive light cone component
of the residual momentum $k$. The relation between the two functions
$S$ and $f$ is given by
\begin{equation}
S(y) = \frac{1}{m_Q} \int\limits_{-m_Q (1-y)}^{\bar\Lambda} dk_+ f(k_+) ,
\end{equation}
from which we infer that the $n^{th}$ moment of the endpoint region
is given in terms of the matrix element $\langle B(v)|
\bar h_v (i D_+)^n h_v |B(v)\rangle $.

The function $f$ is a universal distribution function, which
appears in all heavy-to-light inclusive decays; another example
is the decay $B \to X_s \gamma$ \cite{NeubertBsg,BiMotion},
where this function determines
the photon-energy spectrum in a region of order $1/m_Q$ around
the $K^*$ peak.

In principle $f$ has to be determined by other methods than the $1/m_Q$
expansion, e.g.\ from lattice calculations or from a model, or it has
to be determined from experiment by measuring the photon spectrum
in $B \to X_s \gamma$ or the lepton spectrum in $B \to X_u \ell \bar{\nu}$.
In the context of the model ACCMM model \cite{ACCMM} $f$ has been
calculated in~\cite{BigiACCM}.

\begin{figure}[t]
   \epsfysize=6cm
   \centerline{\epsffile{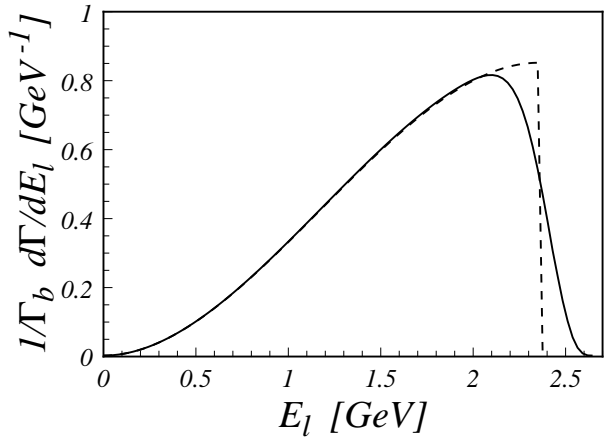}}
\vskip -3mm
   \caption{Charged-lepton spectrum in $B\to X_u \ell  \bar\nu$
decays. The solid line is (\protect{\ref{shape}}) with the ansatz
(\protect{\ref{ftoy}}), the dashed
line shows the prediction of the free-quark decay model. The figure
is from \protect\cite{ManNeu}.}
\label{fig2}
\end{figure}

Some of the properties of $f$ are known. Its support is
$-\infty < k_+ < \bar\Lambda$, it is normalized to unity, and its
first moment vanishes. Its second moment is given by $a_1$, and its
third moment has been estimated \cite{BiMotion,Mzerec}. A one-parameter
model for $f$ has been suggested in \cite{ManNeu}, which incorporates
the known features of $f$
\begin{equation} \label{ftoy}
   f(k_+) = {32\over\pi^2\bar\Lambda}\,(1-x)^2
   \exp\bigg\{ - {4\over\pi}\,(1-x)^2 \bigg\}\,
   \Theta(1-x)  ,
\end{equation}
where $x = k_+ / \bar\Lambda$, and
the choice $\bar\Lambda = 570$ MeV yields reasonable values
for the moments. In fig.~\ref{fig2} we show the spectrum for
$B \to X_u \ell \nu_\ell$ using the ansatz (\ref{ftoy}).

Including the non-perturbative effects yields a reasonably behaved
spectrum in the endpoint region and the $\delta$-function-like
singularities have disappeared. Furthermore, the spectrum now extends
beyond the parton model endpoint;
it is shifted from $E_\ell^{max} = m_Q / 2$ to the physical
endpoint $E_\ell^{max} = M_H / 2$, since $f$ is non-vanishing for
positive values of $k_+ < \bar\Lambda = M_H - m_Q$.

\section{Conclusions and Open Questions}
The $1/m_Q$ expansion obtained from the OPE and HQET offers the unique
possibility to calculate the transition rates for inclusive decays in a QCD
based and model independent framework. The leading term of this expansion
is always the free quark decay, and the first non-trivial corrections are
in general given by the mean kinetic energy of the heavy quark inside the
heavy hadron $\lambda_1$ and the matrix element $\lambda_2$ of the
chromomagnetic moment operator.

The method also allows us to calculate differential distributions, such as
the charged lepton energy spectrum in inclusive semileptonic decays of heavy
hadrons. For this case, the expansion parameter is the inverse of the energy
release $m_b - 2 E_\ell$, where $E_\ell$ is the lepton energy. Close to the
endpoint, the energy  release is small and thus the expansion in its
inverse powers becomes useless. In this kinematic region one may partially
resume the $1/m_Q$ expansion, obtaining a result closely analogous to the
leading twist term in deep inelastic scattering. Particularly in the endpoint
region a non-perturbative function is needed which corresponds to the parton
distributions parametrizing the deep inelastic scattering.

The main focus of this review have been the non-perturbative corrections,
and we did not consider the perturbative ones. In general, the
non-perturbative contributions are typically a few percent, while the
perturbative ones are usually in excess of ten percent. An exception is
the endpoint region of semileptonic decay spectra, where both the perturbative
and the non-perturbative corrections become large and it becomes hard to
disentangle the two.

The method of the $1/m_Q$ expansion allows us to systematically calculate
even purely hadronic {\em inclusive} rates. This is remarkable, since
calculations
of {\em exclusive} hadronic processes are still very model dependent and thus
in
general not reliable. Having control over the purely hadronic inclusive widths
means that one is in the position to calculate lifetimes and branching
fractions
within the $1/m_Q$ expansion.

The leading term of the $1/m_Q$ expansion is the free quark decay, and it is
known that the semileptonic branching fraction calculated in the parton model
is too large by a few percent. The first non-perturbative corrections turn out
to be much too small to explain the low semileptonic branching fraction. The
perturbative corrections are larger and indeed tend to lower the semileptonic
branching fraction, but the effect is still too small to explain the data.
Recent
calculations \cite{Ball,Ballnew} indicate that there are some effects
originating
from the finite mass of the final state quarks, in particular in the channel
$b \to c\bar{c}s$. However, enlarging this channel relative to $b \to
c\bar{u}d$
will increase the average number $n_c$ of charm quarks produced per $b$ decay.
Experimentally, $n_c$ is close to unity, $n_c = 1.04 \pm 0.07$ \cite{GlasRap},
while an explanation of the low semileptonic branching fraction via an
enhancement
of the channel $b \to c\bar{c}s$ would lead to $n_c \sim 1.3$.

The problem of the semileptonic branching fraction is at the level of a two
standard deviation discrepancy, and another problem of about the same
significance
are the lifetimes of $b$ hadrons. Lifetime differences
between $B$ mesons should show up a the level of the $1/m_b^3$ corrections and
thus should be small. This is supported by data; however, the situation is
different
for the lifetime of $b$ baryons. Their lifetime may differ at the $1/m_b^2$
level and hence are expected to be of the order of a few percent.
Experimentally
one finds a large lifetime difference between the $\Lambda_b$ and the $B$
mesons:
$\tau (\Lambda _b) / \tau (B_d) = 0.72 \pm 0.11 $.

If with new and improved data these two problems persist, they will need
clarification
and perhaps will lead to new insights into the strong interaction
aspects of heavy flavour weak decays.


{\small
\begin{thebibliography}{9}
\bibitem{shifman} {M.~Voloshin and M.~Shifman,
               Sov. J. Nucl. Phys. {\bf 45} (1987) 292  and
               {\bf 47} (1988) 511;
               E.~Eichten and B.~Hill,
               Phys.\ Lett.\ {\bf B234} (1990) 511;
               a more complete set of
               references can be found in one of the
               reviews \cite{reviews}.}
\bibitem{isgur/wise}{N. Isgur and M. Wise,
               Phys. Lett. {\bf B232} (1989) 113 and
               {\bf B237} (1990) 527;
               B. Grinstein,
               Nucl. Phys. {\bf B339} (1990) 253;
               H. Georgi,
               Phys. Lett. {\bf B240} (1990) 447;
               A. Falk, H. Georgi, B. Grinstein and M. Wise,
               Nucl. Phys. {\bf B343} (1990) 1.}
\bibitem{reviews}{The subject of HQET is reviewed in:
               H.~Georgi: contribution to the
               {\it Proceedings of TASI--91},
               by R.~K.~Ellis et al. (eds.)
               (World Scientific, Singapore, 1991);
               B.~Grinstein: contribution to {\it High Energy Phenomenology},
               R.~Huerta and M.~A.~Peres (eds.)
              (World Scientific, Singapore, 1991);
               N.~Isgur and M.~Wise: contribution to {\it Heavy Flavors},
               A.~Buras and M.~Lindner (eds.)
              (World Scientific, Singapore, 1992);
               M.~Neubert, Phys.\ Rept.\ {\bf 245} (1994) 259;
               T.~Mannel,  contribution to {\it QCD--20 years later},
               P.~Zerwas and H.~Kastrup (eds.)
              (World Scientific, Singapore 1993).}
\bibitem{russnl}{M. Shifman and M. Voloshin,
                 Sov. J. Nucl. Phys. {\bf 41} (1985) 120;
                 V. Khoze et al.,
                 Sov. J. Nucl. Phys. {\bf 46} (1987) 112.}
\bibitem{CCGincsl}{J.~Chay, H.~Georgi and B.~Grinstein,
                   Phys. Lett. {\bf B247} (1990) 399.}
\bibitem{Bigiincsl}{I. Bigi, N. Uraltsev and A.~Vainshtein,
                    Phys.\ Lett.\ {\bf B293} (1992) 430;
                    I. Bigi et al., Minnesota TPI-MINN-92/67-T (1992) and
                    Phys. Rev. Lett. {\bf 71} (1993) 496.}
\bibitem{Blokincsl}{B.~Blok et al.,
                    Phys. Rev. {\bf D49} (1994) 3356.}
\bibitem{mwincsl}{A.~Manohar and M.~Wise,
                   Phys. Rev. {\bf D49} (1994) 1310.}
\bibitem{tmincsl}{T.~Mannel, Nucl.\ Phys. {\bf B423} (1994) 396.}
\bibitem{FLS93}{A.~Falk, M.~Luke and M.~Savage,
                Phys.\ Rev.\ {\bf D49} (1994) 3367.}
\bibitem{NeubertShape}{M. Neubert, Phys. Rev. {\bf D49} (1994) 2472.}
\bibitem{BiMotion}{I. Bigi et al., Int.\ J. Mod.\ Phys.\ {\bf A9}
                 (1994) 2467.}
\bibitem{ManNeu}{T. Mannel and M. Neubert, preprint CERN-TH.7156/94
                (1994).}
\bibitem{NeubertBsg}{M. Neubert,
                     Phys.\ Rev.\ {\bf D49} (1994) 4623.}
\bibitem{C12}{F. Gilman and M. Wise,
             Phys. Rev. {\bf D20} (1979) 2392;
             A. Buras et al.,
             Nucl. Phys. {\bf B370} (1992) 69.}
\bibitem{C7}{B. Grinstein, R. Springer and M. Wise,
             Nucl. Phys. {\bf B319} (1988) 271;
             M. Misiak,
             Nucl. Phys. {\bf B393} (1993) 23.}
\bibitem{Luke}{M. Luke,
               Phys. Lett. {\bf B252} (1990) 447.}
\bibitem{Mzerec}{T. Mannel, Phys. Rev. {\bf D50} (1994) 428. }
\bibitem{Bisumrule}{I. Bigi et al.,
                    preprint CERN-TH-7250/94 (1994),
                    hep-ph/9405410.}
\bibitem{BBsumrule}{P. Ball and V. Braun,
               Phys. Rev. {\bf D49} (1994) 2472.}
\bibitem{QCDradcorr}{A. Ali and E. Pietarinen,
                    Nucl. Phys. {\bf B154} (1979) 519;
                    N. Cabibbo, G. Corbo and L. Maiani,
                    Nucl. Phys. {\bf B155} (1979) 83;
                    G. Corbo,
                    Nucl. Phys. {\bf B212} (1983) 99;
                    M. Jezabek and J. H. K\"uhn,
                    Nucl. Phys. {\bf B320} (1989) 20;
                    A. Falk et al.,
                    Phys. Rev. {\bf D49} (1994) 3367.}
\bibitem{ACCMM}{G. Altarelli et al.,
                Nucl. Phys. {\bf B208} (1982) 365.}
\bibitem{Paschos}{E. Paschos, contribution to this conference,
                  C. Jin, W. Palmer and E. Paschos,
                  Phys.\ Lett.\ {\bf B329} (1994) 364.}
\bibitem{BigiACCM}{I. Bigi et al,
                   preprint CERN-TH.7159/94. }
\bibitem{Ball}{E. Bagan et al., Phys. Lett. {\bf B342} (1995) 362.}
\bibitem{Ballnew}{E. Bagan et al.,
                  preprint CERN-TH.95--25.}
\bibitem{GlasRap}{P. Roudeau, Rapporteur
                  talk at the ICHEP 94 Conference, Glasgow, Scottland.}
\end{thebibliography}}
\end{document}